\documentclass[usenatbib,useAMS]{mnras}

\DeclareRobustCommand{\VAN}[3]{#2}
\let\VANthebibliography\thebibliography
\def\thebibliography{\DeclareRobustCommand{\VAN}[3]{##3}\VANthebibliography}

\usepackage{graphicx}	
\usepackage{amsmath}	
\usepackage{amssymb}	
\usepackage{amsfonts}
\usepackage{acronym}
\usepackage{bm}
\usepackage{subfig}
\usepackage{physics}

\usepackage[utf8]{inputenc}
\usepackage[T1]{fontenc}
\usepackage{ae,aecompl}

\usepackage{newtxtext,newtxmath}

\newcommand{\Nobs}{\ensuremath{N_\mathrm{obs}}}
\newcommand{\Npix}{\ensuremath{N_\mathrm{pix}}}
\newcommand{\dL}{\ensuremath{d_\mathrm{L}}}

\title[Are BBH mergers isotropically distributed?]
{Are stellar--mass binary black hole mergers isotropically distributed?}

\author[Stiskalek et al.]
{Richard Stiskalek$^{1}$\thanks{\href{mailto:richard.stiskalek@protonmail.com}{richard.stiskalek@protonmail.com}},
John Veitch$^{1}$, \& Chris Messenger$^{1}$
\\
$^{1}$Institute for Gravitational Research, University of Glasgow, Glasgow, G12 8QQ, United Kingdom}

\date{Accepted XXX. Received YYY; in original form ZZZ}

\pubyear{2020}

\begin{document}
\label{firstpage}
\pagerange{\pageref{firstpage}--\pageref{lastpage}}
\maketitle

\definecolor{debianred}{rgb}{0.84, 0.04, 0.33}
\acrodef{BBH}{binary black hole}
\acrodef{PSD}{power spectral density}
\acrodef{PDF}{probability density function}

\begin{abstract}
The Advanced LIGO and Advanced Virgo gravitational wave detectors have detected
a population of binary black hole mergers in their first two observing runs. For
each of these events we have been able to associate a potential sky location
region represented as a probability distribution on the sky. Thus, at this point
we may begin to ask the question of whether this distribution agrees with the
isotropic model of the Universe, or if there is any evidence of anisotropy. We
perform Bayesian model selection between an isotropic and a simple anisotropic
model, taking into account the anisotropic selection function caused by the
underlying antenna patterns and sensitivity of the interferometers over the
sidereal day. We find an inconclusive Bayes factor of $1.3:1$, suggesting that
the data from the first two observing runs is insufficient to pick a preferred
model. However, the first detections were mostly poorly localised
in the sky (before the Advanced Virgo joined the network), spanning large portions
of the sky and hampering detection of potential anisotropy. It will be
appropriate to repeat this analysis with events from the recent third LIGO
observational run and a more sophisticated cosmological model.
\end{abstract}




\section{Introduction}
\label{sec:introduction}
%
The first detection of gravitational waves by Advanced LIGO~\citep{Aasi:2013wya,
aLIGO, Harry:2010zz, GW150914}, revealed the existence of a detectable
population of coalescing stellar--mass \acp{BBH}. This was confirmed by the
subsequent \ac{BBH} detections in the first two observing runs
(~\cite{O1O2catalog}, O1~\citep{O1BBH}, O2~\citep{GW170104,GW170608,GW170814}),
during the latter of which the Advanced Virgo detector joined the
network~\citep{AdVirgo}. The location of the mergers can be determined by
performing a coherent analysis of the data from the two-or three-detector
network, using either a rapid localisation algorithm~\citep{BAYESTAR} or a full
parameter estimation method~\citep{LALInference}. Although the initial
detections could be constrained to only tens to hundreds of square degrees, the
addition of Advanced Virgo to the network has resulted in improved localisation
of subsequent detections such as GW170814~\citep{GW170814}.

With $10$ \ac{BBH} detections being announced to date from O1 and O2, it is
possible to begin to determine the properties of the source population, such
as the rate, sky and mass
distribution~\citep{GW150914:RATES,O1BBH,O2populations}. This type of question
is addressed by a hierarchical analysis of the sources, which must include the
effect of the detector sensitivity on the detectable events. Previous studies
have looked at the variation of the selection function with mass, spin, and
sidereal time~\citep{PhysRevD.82.104006,2015ApJ...806..263D,
Ng:2018neg,Chen:2016luc}. The situation is further complicated by the large
uncertainties on the source location, particularly during O1 when only the two
LIGO detectors were operational.

Standard cosmological models are consistent with the cosmological principle,
that the properties of the Universe are the same for all observers when viewed
on large scales. One of the two testable consequences of this is that the 
matter distribution of the Universe, and as an extension gravitational wave sources,
would be distributed isotropically. Analysis of the cosmic microwave background
temperature and polarisation fluctuations using Planck observations have concluded
that anisotropy is strongly disfavoured~\citep{2016PhRvL.117m1302S}. However,
gravitational wave observations provide an independent channel through which to
verify this conclusion. The largest observed structure (defined by the spatial
distribution of gamma-ray burst events) is $\sim \mathrm{Gpc}$ in
size~\citep{2013arXiv1311.1104H} and corresponds to the
currently most distant detected \ac{BBH} events. Hence, it will be interesting
to study whether the observed sky distribution of gravitational wave events
matches that of the local structure seen via electromagnetic channels.

In this work we address the issue of the distribution of sources over the sky,
taking into account the sky-variation of the selection function of the detector
network during the first two observing runs. Our intent is to compare two
models; an isotropic source population and an anisotropic model that divides the
sky into a finite set of pixels.

In Section~\ref{sec:analysis} we describe these
models together with our analysis, report the results in
Section~\ref{sec:results} and in Section~\ref{sec:conclusion} we summarise the
results, discuss the model and its astrophysical significance.

\subsection{Data}
\begin{figure*}
    \centering
    \includegraphics[width=0.8\textwidth]{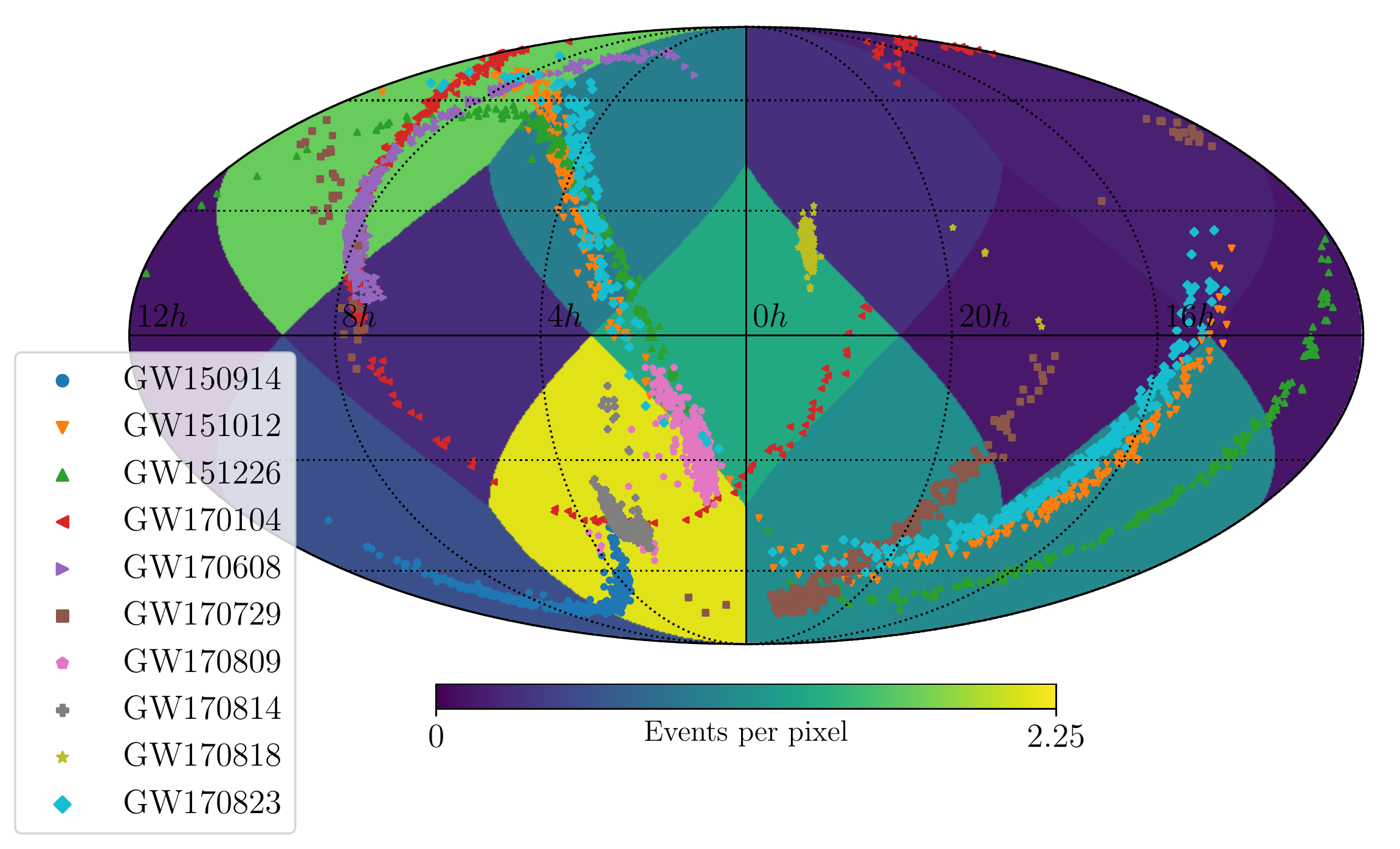}
    \caption{
    A sky map divided into $12$ pixels of equal area. The colour scale indicates
    the number of events per pixel (calculated by counting the number of
    posterior samples in each pixel, normalised by the total number of samples
    per event). About $\sim~2.2$ events are assigned to the highest density
    pixel as GW170814 and partially GW170809 are tightly localised to that
    pixel. Some pixels contain almost no samples; the pixel with the largest
    fraction of events contains $\sim~15$ times more samples than the pixel with
    the smallest fraction.}
    \label{fig:eventLocalisation}
\end{figure*}

The posterior samples containing information about the sky localisation of
events were taken from the LIGO data releases via the Gravitational-Wave Open
Science Center (GWOSC)~\citep{Vallisneri:2014vxa}, and the following events were
used: GW150914, LVT151012, GW151226, GW170104, GW170608, GW170729, GW170809,
GW170814, GW170818 and GW170823~\citep{O1O2catalog}. In the subsequent analysis
we limit each event to $5,000$ posterior samples to reduce the computational
complexity and weight each event equally. Figure~\ref{fig:eventLocalisation}
shows a scatter plot of samples from the sky posterior probability densities for
all events, binned into $12$ pixels. A large proportion of all samples is coming
from a single region of the sky (mostly due to GW170814 and GW170809 being
tightly localised), whereas some areas of the sky have almost no samples. As for
the \ac{PSD} curves, we separately consider the \ac{PSD} estimates at the time
of event~\citep{PSDevent} and run-averaged~\citep{PSDO1H1, PSDO1L1, PSDO2}
estimates, using the publicly available noise curves~\citep{O1O2catalog}.

\section{Analysis}\label{sec:analysis}
The focus of this work is to represent the distribution of \ac{BBH} sources on
the sky. To do so, there are two readily available methods of decomposing the
sky into either a finite set of pixels or spherical harmonics. In this work, for
simplicity, we opt for a model that decomposes the sky into a finite set of
pixels.

In this pixelated, anisotropic model we divide the sky into $\Npix$ pixels of
equal area $4\pi / \Npix $ steradians. Each pixel $i$ has a parameter $a_i$,
which is further referred to as a pixel weight. This pixel weight describes the
stellar--mass \ac{BBH} merger rate per steradian, per unit time, per unit
volume within that pixel, i.e. the units of  $\bm{a}$ are 
$\left[\bm{a}\right] = \mathrm{Gpc}^{-3}\mathrm{yr}^{-1}\mathrm{sr}^{-1}$,
and both models assume homogeneous distribution of mergers within each pixel.
Given that the pixel weights give the rate per steradian, we also infer the
total rate $R$ ($\left[R\right] = \mathrm{Gpc}^{-3}\mathrm{yr}^{-1}$) as a
sum over the whole sky,
\begin{align}\label{eq:totalRate}
    R=\Delta\Omega \sum_{i}^{\Npix} a_i,
\end{align}
i.e. the sum of the product of pixel weights and pixel areas $\Delta\Omega$,
making $R$ conditionally dependant on $\bm{a}$.

In case of the anisotropic model the number of pixels was chosen to be
$N_{\mathrm{ani}}=12$, since in the subsequent analysis we work with HEALPix
maps~\citep{2005ApJ...622..759G} and the Python package Healpy where $12$ is
the minimum number of pixels available, making the analysis computationally
inexpensive, while still ensuring the model has freedom to detect large-scale
anisotropy. For the isotropic model we simply use $N_{\mathrm{iso}}=1$, wherein
the whole sky is described by a single pixel, forcing a uniform distribution
of the rate of \ac{BBH} mergers. 

HEALPix divides the sky into $N$ pixels with a fixed distribution.
Therefore to allow for different angular distribution of pixels we introduce
three Euler angles  $\bm{\phi}\equiv \{\alpha, \beta, \gamma\}$ to describe the
reference position and orientation of the HEALPix grid. The Euler angles
$\bm{\phi}$ are used to calculate a rotation matrix to rotate the GW datasets
to accommodate for different orientations of the fixed HEALpix grid.

We perform a Bayesian analysis to estimate the pixel weights
$\bm{a}\equiv\{a_i\}$ and to perform model selection between the isotropic and
anisotropic models. For a given model $I\in(\text{ISO},\text{ANISO})$,
gravitational wave data sets $\{x_j\}$ for each of a number of detections
$\Nobs$, and where $\bm{\mathrm{D}}$ indicates detection (described in
Subsection~\ref{subsec:probDet}), the posterior on the parameters $\bm{\theta} =
\{\bm{a},\,\bm{\phi}\}$, where $\bm{a}$ are the pixel weights and $\bm{\phi}$
the Euler angles, is given by
\begin{equation}\label{eq:posterior}
    \begin{split}
	p\left(\bm{\theta} \mid \Nobs,\{x_j\}, \bm{\mathrm{D}}, I\right)
    &= \frac{p\left(\Nobs, \{x_j\} \mid \bm{\theta}, \bm{\mathrm{D}}, I)\right)
    p\left(\bm{\theta}\mid \bm{\mathrm{D}}, I\right)}
    {p\left(\Nobs,\{x_j\}\mid \bm{\mathrm{D}}, I\right)}.
    \end{split}
\end{equation}

Having obtained the expression for the posterior (Eq.~\ref{eq:posterior}) we
calculate the evidences for each model,
\begin{align}\label{eq:evidence}
	p\left(\Nobs, \{x_j\} \mid \bm{\mathrm{D}}, I\right)
	=
	\int p\left(\Nobs, \{x_j\},\mid \bm{\theta}, \bm{\mathrm{D}}, I\right)
	p(\bm{\theta} \mid \bm{\mathrm{D}}, I)d^{n}\bm{\theta}.
\end{align}
The ratio of evidences from the two models gives the Bayes factor, which
indicates the relative support for one over the other that is imparted by
a particular set of observations. For the isotropic model there is only pixel,
hence one pixel weight $a_{\text{iso}}$ related to the total rate as $R=4\pi
a_{\text{iso}}$. Moreover, the isotropic model is also independent of the Euler
angles $\bm{\theta}$. We use the nested sampling algorithm~\citep{skilling2006}
in a Python package CPNest~\citep{john_veitch_2017_835874} to sample
the posterior and obtain the evidence for the anisotropic model. The isotropic model
is evaluated analytically.

\subsection{Selection function}\label{subsec:probDet}
The selection function describes the interferometer sensitivity for a given
distribution of \ac{BBH} mergers as a function of sky position $\Omega$ and
luminosity distance $d_\mathrm{L}$. We consider both the O1 and O2 runs
separately and define the selection function as
\begin{equation}\label{eq:selection}
    p\left(\bm{\mathrm{D}} \mid \Omega, d_{\text{L}}, I\right)
    =
    \int\limits_{\rho_{\mathrm{thresh}}}^{\infty} d\rho
    p\left(\rho \mid \Omega,d_{\text{L}}, \bm{\vartheta}, I\right)
    p(\bm{\vartheta}|\Omega,I)p(d_{\text{L}}|I),
\end{equation}
where a signal is modelled as detected if its signal-to-noise-ratio $\rho$ is
measured above a predefined threshold~\citep{H0paper}. The probability of
detection, conditional on the sky location, is computed by marginalising over
the prior ranges on the distance and the remaining source parameters denoted
by $\bm{\vartheta}$. The mass distribution of the primary is
assumed to be a power-law $p\left(m_1\right)\propto m_1^{-2.35}$, with a minimum
mass of $5\,M_\odot$, and the secondary to be uniformly distributed, being
always less massive than the primary. A constraint is placed on the
sum of the primary and secondary to be always less than $100\,M_\odot$. This
choice of BBH mass distribution is consistent with the estimated distribution
following the O1 observing run~\citep{O1BBH}.

\begin{figure*}
	\centering
	\subfloat[Mean $\langle p\left( \bm{\mathrm{D}} \mid \Omega, I\right)\rangle$ for the O1 run]
    {\includegraphics[width=0.49\textwidth]{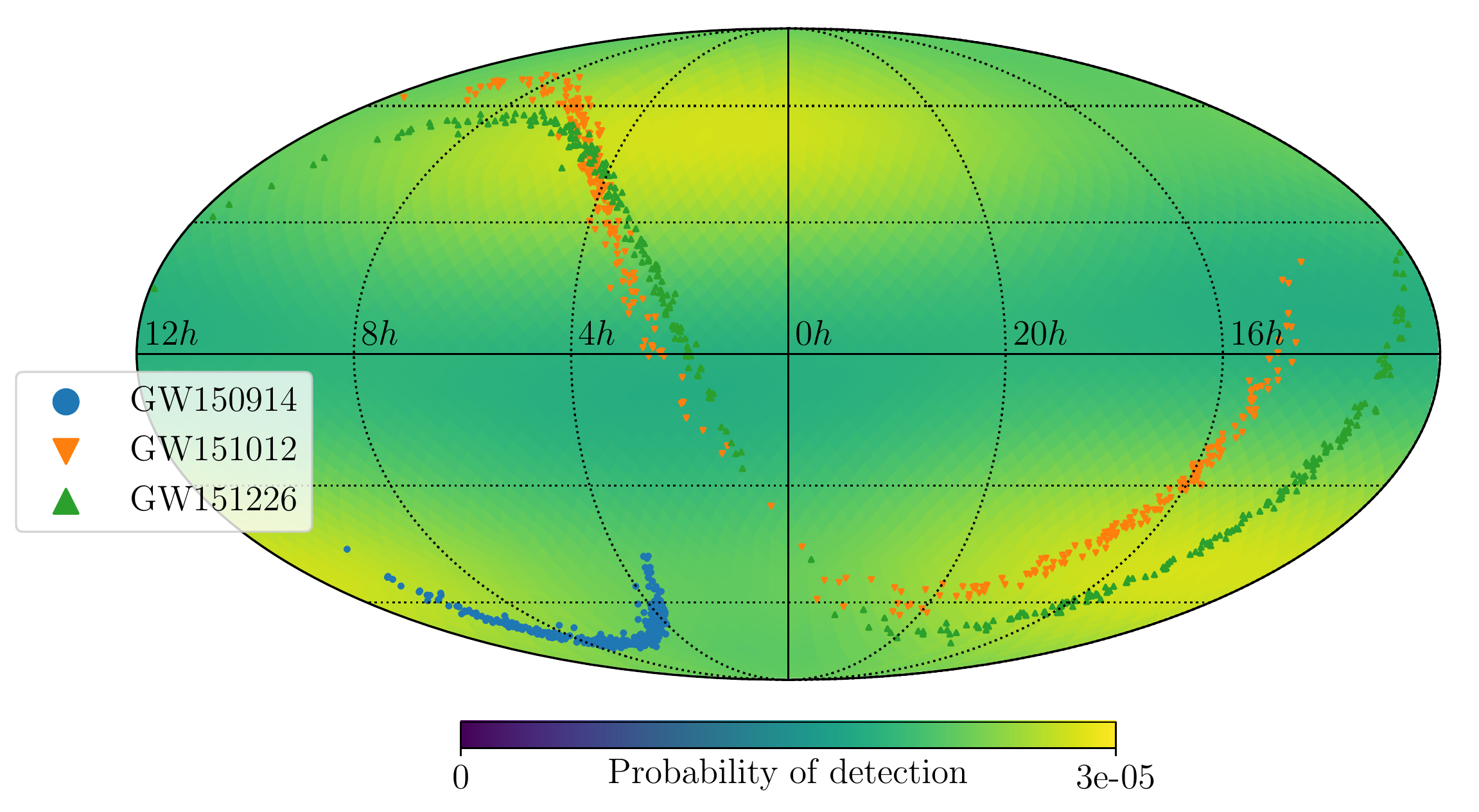}
	\label{fig:pofdO1}}
	\hfill
	\subfloat[Mean $\langle p\left( \bm{\mathrm{D}} \mid \Omega, I\right)\rangle$
	for the O2 run]{\includegraphics[width=0.49\textwidth]{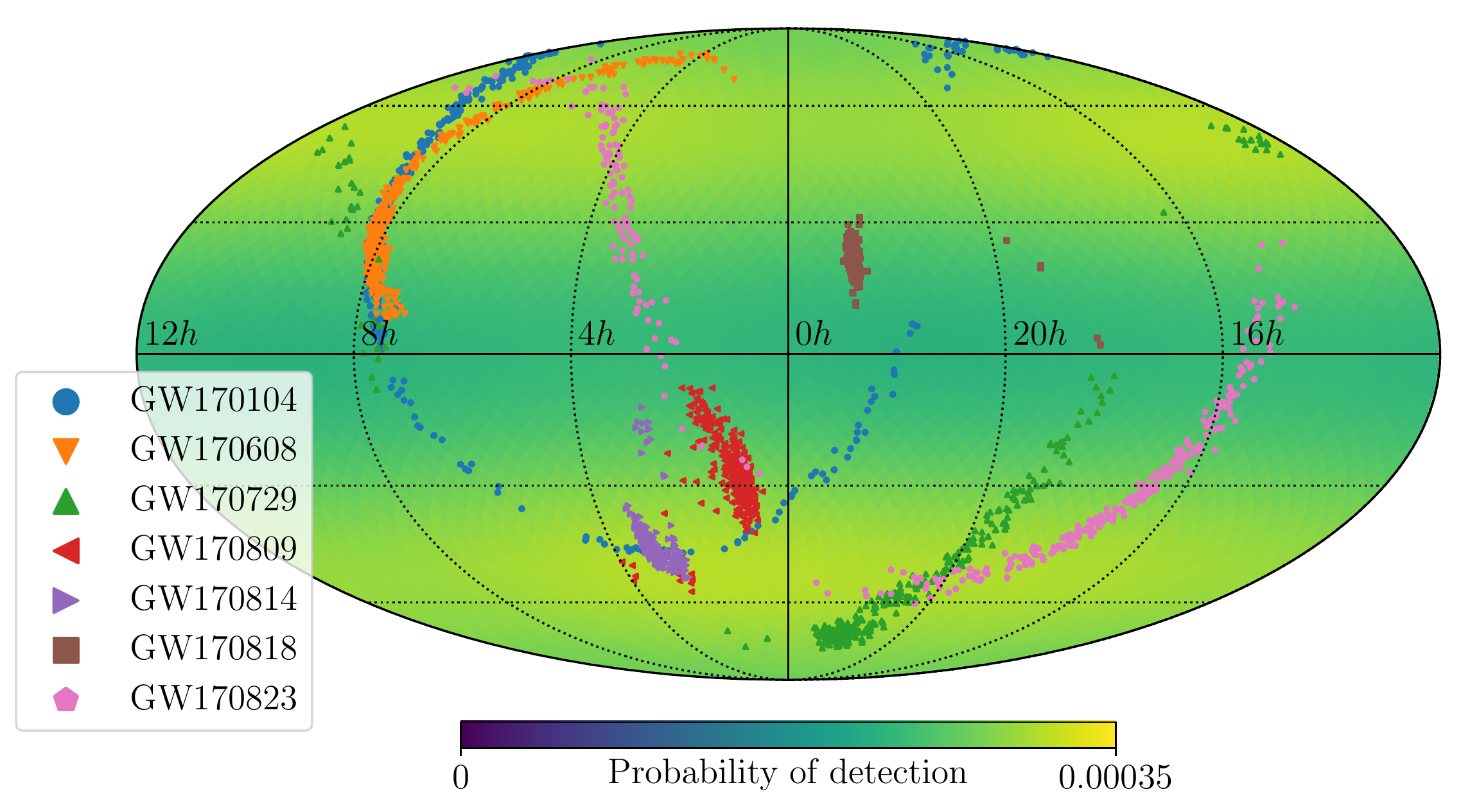}
    \label{fig:pofdO2}}
    \caption{Mean probability of detection maps for the O1 (a) and O2 (b) runs, with
    posterior samples for events from O1 and O2 scattered over. The maps were
    calculated with the average \ac{PSD} for each run, marginalising the
    selection function over distance and time averaging the
    $p\left(\bm{\mathrm{D}} \mid \Omega, t, I\right)$, where $t$ is the GPS
    time, over the times the detectors were operational.}
\label{fig:meanPofd} 
\end{figure*}

Since we are interested in anisotropy, we must consider the directional
sensitivity of the detectors. The detectors are most sensitive to sources
positioned directly above and below the detector~\citep{Anderson:2000yy}. As the
Earth rotates, the antennae response function is smeared out in right-ascension,
but as noted in~\citet{Chen:2016luc}, there is a tendency for the detector
sensitivity to vary over the course of the day due to human activity near the
sites as well. This, therefore, produces a selection function which is specific
to the observing conditions during a particular observing run.
A full treatment of this would need to use the actual noise power spectra as a
function of time throughout the observing run, but we approximate this by
taking the power spectrum at the time of detections and the run-averaged
power spectrum. Integrating the antenna response function over sidereal
time, using a weight of one when the detector was taking science quality
data, we obtain the run-averaged selection function~\citep{Vallisneri:2014vxa}.
We show this averaged selection for both
observing runs in Fig.~\ref{fig:meanPofd}. The ratio of the maximum and minimum
averaged selection function in Fig.~\ref{fig:meanPofd} is about $1.5$ and $1.4$
for O1 and O2, respectively, indicating that in the O2 run the detectors showed
marginally less preference for certain directions. This can be explained by the
increased up-time of the O2 run, which reduces the overall variation in right
ascension.

\begin{figure}
    \centering
    \includegraphics[width=0.9\columnwidth]{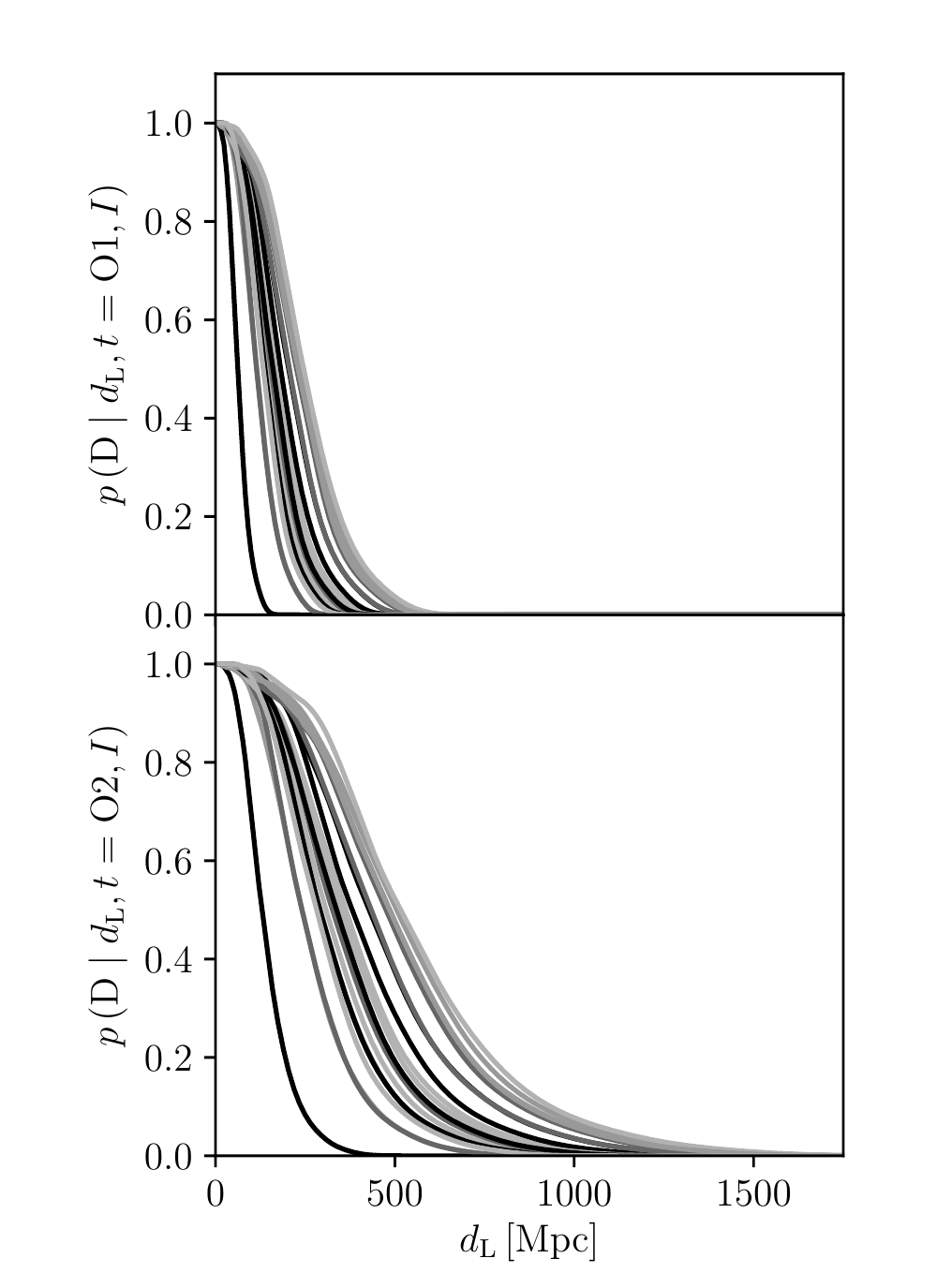}
    \caption{Comparison of survival functions for the O1 (top panel)
    and O2 (bottom panel) runs, where each line is the survival function sampled
    at a discrete
    sky position. The detectors are most sensitive towards directions above and
    below, and least sensitive to directions in the plane of the detector. There
    is a significant increase in sensitivity between the two runs.} 
    \label{fig:survivalPlots}
\end{figure}

Fig.~\ref{fig:survivalPlots} shows the resulting selection functions
(probability of detection) for O1 and O2, plotted as a function of distance for
various sky positions. This shows the effect described above, with anisotropic
sensitivity variations favouring the latitudes directly above the LIGO detectors,
but with the response function smeared out in right ascension by the Earth's rotation.
Nonetheless, there is still some variation in right ascension, which is more
pronounced for the O1 run. Comparing the two panels also shows the overall
increase in detection sensitivity in the O2 run.

\begin{figure*}
  \centering
	\subfloat[$p\left( \bm{\mathrm{D}} \mid \Omega, I\right)$ at the time of
	GW151226.]
    {\includegraphics[width=0.49\textwidth]
	{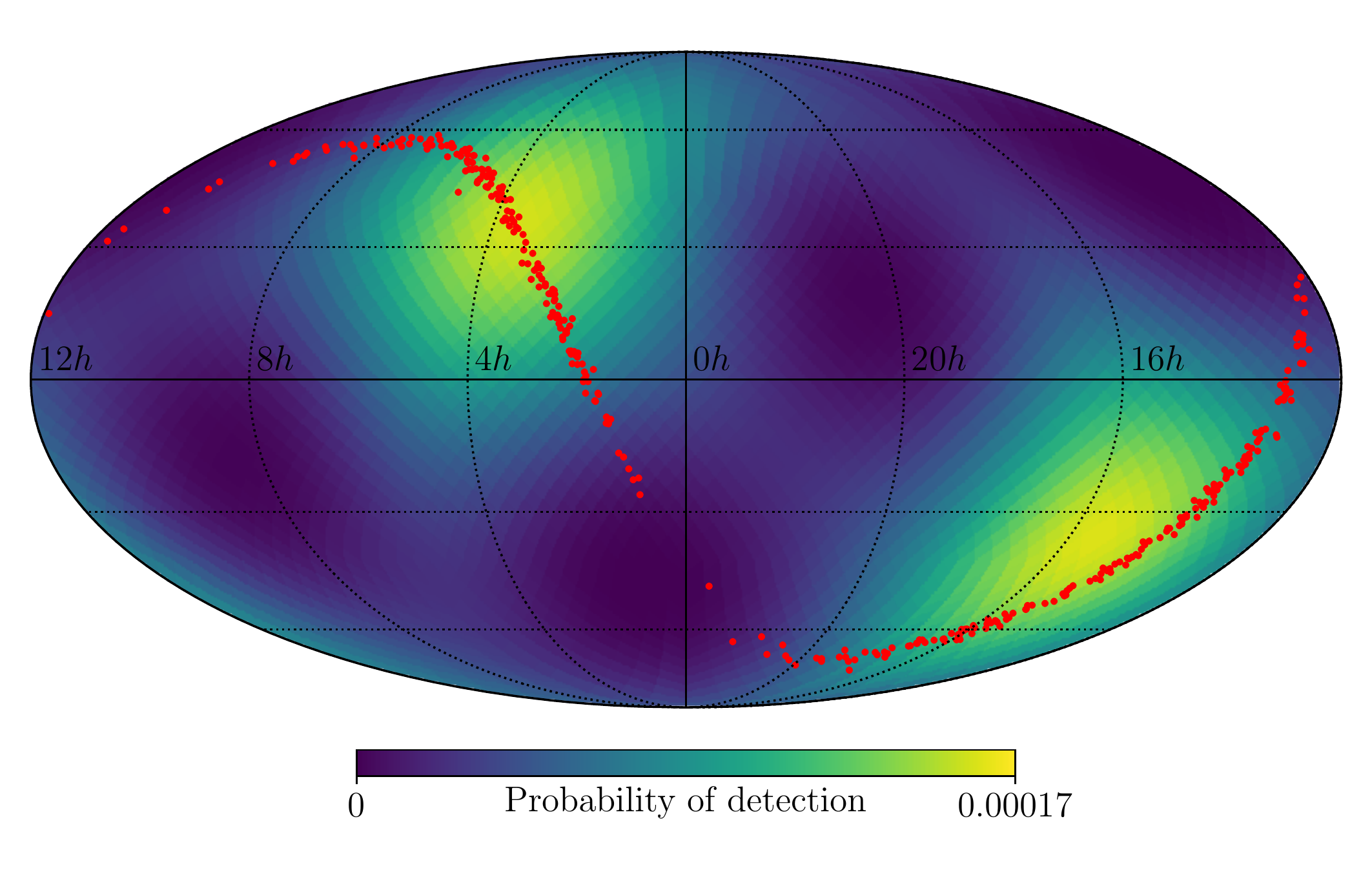}
    \label{fig:pofdGW151226}}
	\hfill
	\subfloat[$p\left( \bm{\mathrm{D}} \mid \Omega, I\right)$ at the time of
    GW170104.]
    {\includegraphics[width=0.49\textwidth]
	{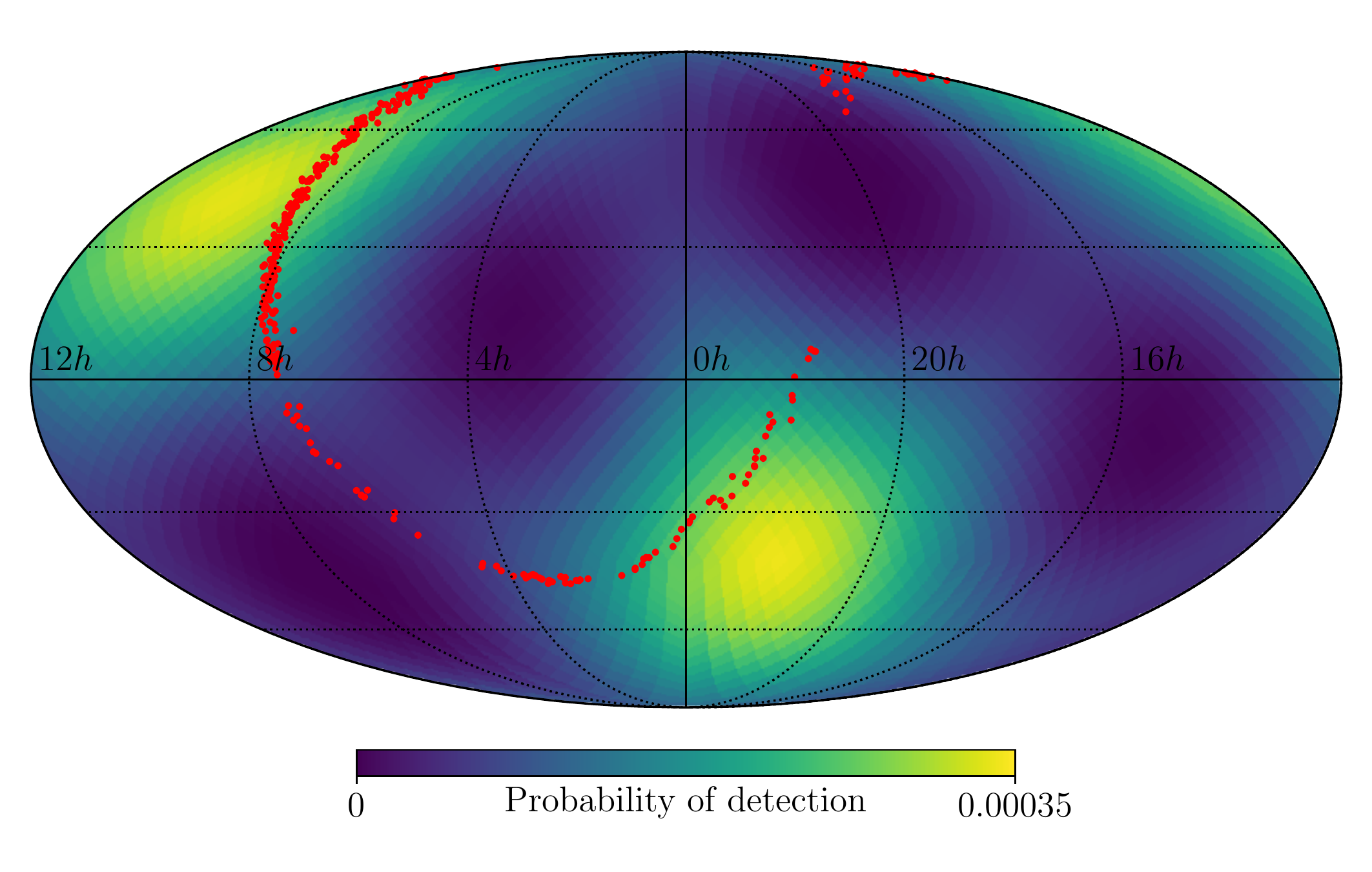}\label{fig:pofdGW170104}}
    \caption{Probability of detection maps at the time of GW151226 (a) from O1
    run and GW170104 from O2 run (b), with the respective posterior samples scattered
    over. The maps were calculated with the \ac{PSD} at the time of event
    detection and marginalising the selection function over luminosity distance.
    While the pattern is the same due to presence of only two LIGO detectors,
    the magnitude of the detection probability increases for GW170104 as the
    detectors were more sensitive during the O2 run.}
	\label{fig:instantProb}
\end{figure*}

Marginalising out the luminosity distance we obtain the probability of detection
as a function of right ascension and declination. To calculate the probability
of detection at the time of events we use the \ac{PSD} estimate at the time of
the \ac{BBH} merger. This, along with the merger time, fully specifies the
orientation of the network geometry in an Earth-fixed coordinate system which
rotates in time along with the Earth. Examples of the probability of detection
function for GW151226 and GW170104 are shown in Fig.~\ref{fig:instantProb}. The
figure also shows the brights spots of high probability from which a detection
would be expected, given the antenna pattern, and the relative increase in
sensitivity (about one order of magnitude) between the two runs.
The probability of detection maps are rendered on a higher resolution map
($3,072$ pixels) compared to the $12$-pixel map of pixel weights, which
describe the \ac{BBH} merger rate in the anisotropic model.

When calculating the probability of detection we only consider
the LIGO Hanford-Livingston detector network since these were
the only detectors used to determine detection in the O1 and O2
runs~\citep{O1O2catalog}. The Advanced Virgo detector joined at the
end of the O2 run and was only used for the subsequent parameter
estimation~\citep{Virgo}, not to determine detections.
Consequently, the probability of detection calculation does not include the
Advanced Virgo. Nevertheless, despite Virgo's lower sensitivity during O2
compared to the LIGO detectors, addition of its data for some of the later O2
events (such as GW170814) yielded significantly better sky localisation
estimates compared to the earlier detections~\citep{GW170814}.

\subsection{Prior}\label{subsec:prior}
In the isotropic model, the only free parameter of the model is $R$, the overall
rate of \ac{BBH} mergers (Eq.~\ref{eq:totalRate}). We choose a Jeffreys prior,
\begin{equation}\label{eq:prioriso}
    p\left(R \mid \mathrm{ISO}\right) \propto R^{-1/2},
\end{equation}
which allows a direct comparison with rate estimates published by the LIGO-Virgo
Collaboration using sophisticated measurements of sensitive time-volume and
models of the source population~\citep{O2populations}.

In the anisotropic model with $\Npix$ equal-area pixels, we have parameters
$a_1, \ldots, a_{\Npix}$. Since we wish to compare it to the isotropic model, we
specify the prior on the overall rate $R$ to have the same Jeffreys form as
Eq.~\ref{eq:prioriso}, so
\begin{equation}
	p\left(R \mid \mathrm{ANISO}\right)
    \propto
    R^{-1/2}
    \propto
	\int p\left(R \mid \bm{a}\right)
	p\left(\bm{a} \mid \mathrm{ANISO}\right) d^{\Npix}\bm{a}.
\end{equation}
Since we do not favour any particular pixel a-priori, the prior should be symmetric
under $a_i\leftrightarrow a_j$. The following prior fulfils this criteria, while being
uniform on the $N-1$-simplex of possible $\bm{a}$\textit{s} for a particular
$R$,
\begin{equation}\label{eq:prioraniso}
	p(\vec{a}|\mathrm{ANISO})
	\propto
	R^{-(2\Npix-1)/2}
	=
	\left( \frac{4\pi}{\Npix} \sum_{i}^{\Npix} a_i\right)^{-(2\Npix-1)/2}.
\end{equation}

\begin{table}
\centering
\begin{tabular}{l l} 
    \hline\hline
    Pixel weight $a_i$ & $\left[0,\,25\right]\,\mathrm{Gpc}^{-3}\mathrm{yr}^{-1}\mathrm{sr}^{-1}$\\
    Total astrophysical rate $R$ & $\left[10^{-5},\,750\right]\mathrm{Gpc}^{-3}\mathrm{yr}^{-1}$
    \\
    First Euler angle about the $z$-axis $\alpha$ & $\left[0,2\pi\right]$
    \\
    Second Euler angle about the $y$-axis $\cos\beta$ & $\left[0.75, 1\right]$
    \\
    Third Euler angle about the $z$-axis $\gamma$ & $\left[0,2\pi\right]$\\
    \hline\hline
\end{tabular}
\caption{Prior ranges on the inferred parameters $\bm{\theta}$ and the total rate $R$.}
\label{table:boundaries}
\end{table}

Boundaries found in Table~\ref{table:boundaries} are applied to the inferred
parameters. Note that the boundary on the total rate $R$ is well outside of the
range found by the O1 and O2 run~\citep{O2populations}. The Euler angles rotate
the angular distribution of pixels, however without restricting the range of
rotations the posterior would become strongly degenerate as multiple
combinations of the Euler angles would correspond to the identical orientation
of the HEALpix grid. Our intent in setting the boundaries for the Euler angles
is twofold. First, we choose a uniform rotation prior over the ranges as shown
in Table~\ref{table:boundaries}. Second, to avoid degeneracy of the likelihood we enforce
a condition wherein no pixel centre can be rotated beyond its original boundary.
An example of this degeneracy is rotating the pixel grid about the $z$-axis by
an angle that corresponds to the angular separation of two nearby pixel centres
that lie in in the $x-y$ plane. The range of $\cos\beta$ was restricted
(Table~\ref{table:boundaries}) as values outside this range would violate the
aforementioned second condition.

\subsection{Likelihood}\label{subsec:likelihood}
In our specific problem, the likelihood $p(\Nobs,\{x_j\} \mid \bm{\theta},I)$
depends on the data for each detected event (yielding estimates of their sky
position, distance, masses and detection time) and the selection function. The
likelihood can be split into the so-called number and event likelihood
respectively,
\begin{equation}\label{eq:likelihood}
	p \left(\Nobs, \{x_j\} \mid \bm{\theta}, \bm{\mathrm{D}}, I\right)
	=
	P\left(\Nobs \mid \bm{\theta},  \bm{\mathrm{D}} , I\right)
	p\left(\{x_j\} \mid \bm{\theta}, \bm{\mathrm{D}}, I\right),
\end{equation}
where $\bm{\mathrm{D}}$ indicates detection and the use of the
interferometers' \acp{PSD}~\citep{2018arXiv180902063M}.

If we divide the observation time into $n$ segments, and assume the probability
of detection to be constant over the segment duration, then we can write the
probability of detecting $\Nobs$ events as a Poisson binomial distribution,
that is
\begin{equation}\label{eq:numberLikelihood}
    \begin{split}
        P\left( \Nobs \mid \bm{\theta}, \bm{D}, I\right)
        &=
        \binom{n}{\Nobs} \prod_{j=1}^{\Nobs}
        p\left(N \geq 1 \mid \bm{\theta}, \bm{\mathrm{D}}_j, I\right)\\
        &\times
        \prod_{k=1}^{n - \Nobs}
        P\left(N = 0 \mid \bm{\theta}, \bm{\mathrm{D}}_k, I\right),
    \end{split}
\end{equation}
so that $\bm{\mathrm{D}}_{j, k}$ is the probability of detection evaluated in the
$j$-th segment. We define success as drawing $1$ or more events within
a segment and a failure as drawing no events,
\begin{equation}
    p\left(N = 0 \mid \bm{\theta}, \bm{\mathrm{D}}_k, I\right)
    =
    \exp\left[-\hat{N}_k\right],
\end{equation}
where $\hat{N}_k$ is the expected number of detections from the $k$-th segment
and 
$p\left(N \geq 1 \mid \bm{\theta}, \bm{\mathrm{D}}_k, I\right)
=
1 - p\left(N = 0 \mid \bm{\theta}, \bm{\mathrm{D}}_k, I\right)$.
In the limit of large $n$ (equivalently $\hat{N}_j \ll 1$),
we may approximate Eq.~\eqref{eq:numberLikelihood} as
\begin{equation}\label{eq:simplified_likelihood}
    P\left( \Nobs \mid \bm{\theta}, \bm{D}, I\right)
    \propto
    \left(\prod_{j=1}^{\Nobs} \hat{N}_j \right)
    \exp \left[ - \sum_{k=1}^{n} \hat{N}_k \right].
\end{equation}

The number of expected detections in the $j$-th time segment of duration
$\mathrm{d}t$ is the convolution of the differential rate and the probability
of detection:
\begin{equation}\label{eq:expected_detections_segment}
    \hat{N}_j = \mathrm{d}t V_\mathrm{L}
    \int
    \dv{R}{\Omega}
    p\left(\bm{\mathrm{D}}_j \mid \Omega,I\right)
    \mathrm{d}\Omega.
\end{equation}
First, because in the anisotropic model we allow the astrophysical merger rate
to vary with sky position, we cannot fully separate it when considering
the number of detections from a given time segment in
Eq.~\eqref{eq:expected_detections_segment}.
Second, for both models we assume homogeneity, therefore $V_{\mathrm{L}}$,
the luminosity volume, can be taken outside the integral in
Eq.~\eqref{eq:expected_detections_segment} as long as
$V_{\mathrm{L}}$ matches the volume over which we marginalised out the
probability of detection function. Because the detection probability
vanishes at sufficiently high distances, our result will be independent of the
choice of $V_{\mathrm{L}}$, provided we consider a volume much larger than the
observable one.

For this analysis we have assumed a simple distance prior, independent of sky
location, $p\left(\dL\right) \propto d_{\text{L}}^{2}$, which is consistent
with a static, Euclidean universe. Similarly, we have also assumed an
underlying astrophysical rate that is constant with distance (and therefore
with redshift). As our primary aim is to test anisotropy, and not homogeneity,
coupled with the fact that the detections are positioned in the local Universe,
we expect that the impact of these assumptions (which are shared by both models)
to be a second order effect. The probability density function of a
sky position $p\left(\Omega \mid \bm{\theta}\right)$ is taken to be uniform
within each pixel, proportional to its weight $a_i$ and time-independent.

An underlying assumption of our models is that the sky can be pixelated
and that the astrophysical rate and probability of detection are uniform over
each pixel (we render the probability of detection on a higher-resolution
basis). Thus, the differential rate $\mathrm{d}R/\mathrm{d}\Omega$
is equal to the pixel's weight, and
Eq.~\eqref{eq:expected_detections_segment} simplifies to
\begin{equation}
    \hat{N}_j = V_{\mathrm{L}}\frac{T}{n}
    \sum_{k=1}^{\Npix} a_k\,
    p\left(\bm{\mathrm{D}}_j \mid \Omega_k, I \right)
    \Delta\Omega,
\end{equation}
where $a_k$, $\Omega_k$, and $\Delta\Omega$ indicate the $k$-th pixel's weight,
sky location, and angular area, respectively.
We also set the segment duration to be $\mathrm{d}t = T/n$, with $T$ being
the observing run duration and $n$ the number of segments. 

In Eq.~\eqref{eq:simplified_likelihood} we require the sum of the expected
number of detections over all segments: $\hat{N} = \sum_{j=1}^{n} \hat{N}_j$.
Since we assume the pixel weights to be time-independent, the sum over
individual segments then simplifies to
\begin{equation}
    \hat{N}
    =
    V_{\mathrm{L}} T \sum_{k=1}^{\Npix} a_k\,
    \langle p\left(\bm{\mathrm{D}} \mid \Omega_k, I \right) \rangle
    \Delta\Omega,
\end{equation}
where $\langle p\left(\bm{\mathrm{D}} \mid \Omega_k, I \right) \rangle
=
1 / n \sum_{j=1}^{n} p\left(\bm{\mathrm{D}}_j \mid \Omega_k, I \right)$
is the probability of detection as a function of sky position averaged over the
times when the interferometers were operating. The expected number of events is
evaluated separately for O1 and O2, as the selection function and observation
times differ for each run.

The event likelihood in Eq.~\eqref{eq:likelihood} can be split into a product
over all events, assuming they are independent of each other,
\begin{equation}\nonumber
		p\left(\{x_{j}\} \mid \bm{\theta}, \bm{\mathrm{D}}_j,I \right)
		=
		\prod_{j=1}^{\Nobs}
		p\left(x_{j} \mid \bm{\theta}, \bm{\mathrm{D}}_j,I \right).
\end{equation}
This can be further expanded using the Bayes's theorem, noting that the
likelihood $p\left(\bm{\mathrm{D}}_{j} \mid x_j, \bm{\theta}, I\right) = 1$ for
detections:
\begin{equation}\label{eq:eventLikehood}
    \begin{split}
    p\left(x_{j} \mid \bm{\theta},\bm{\mathrm{D}}_j, I\right)
	&=
	\frac{p\left(x_{j} \mid \bm{\theta}, I\right)} {p\left(\bm{\mathrm{D}}_{j} \mid \bm{\theta},I \right)}\\
	&=
	\frac{\iint \mathrm{d}\Omega \mathrm{d}\dL
	p\left(x_j \mid \Omega, \dL, I\right)
	p\left(\Omega, \dL \mid \bm{\theta}, I\right)}
	{\int \mathrm{d}\Omega
	p\left(\bm{\mathrm{D}}_j \mid \Omega, I\right)
        p\left(\Omega, \mid \bm{\theta}, I\right)}.
    \end{split}
\end{equation}
We expand both the prior (numerator) and the evidence (denominator) on the
first line of Eq.~\eqref{eq:eventLikehood} using the marginalisation rule.
Furthermore, in Eq.~\eqref{eq:eventLikehood} the prior can be Monte Carlo
approximated over the GWOSC posterior samples (indexed by $i$). The GWOSC
posterior samples assume a uniform prior over component masses, whereas in our
probability of detection calculation we assumed a power-law distribution.
To correct for this, we add our component mass prior to the GWOSC samples. The
evidence in Eq.~\eqref{eq:eventLikehood} does not depend on the GWOSC samples,
only on the run-averaged probability of detection and so can be marginalised
over distance. With this in mind the final approximate expression for
Eq.~\eqref{eq:eventLikehood} becomes
\begin{equation}\label{eq:final_eventlikelihood}
    p\left(x_{j} \mid \bm{\theta},\bm{\mathrm{D}}_j, I\right)
    \approx
    \frac
    {\sum_{i=1}^{N_j}
    p\left(\Omega_{j}^{i} d_{\mathrm{L}, j}^{i} \mid \bm{\theta}, I\right)
    p\left(m_{1, j}^{i}, m_{2, j}^{i} \mid \alpha \right)/N_j
    }
    {
    \sum_{k=1}^{\Npix} \Delta \Omega
    p\left(\bm{\mathrm{D}}_j \mid \Omega_k, I\right)
    p\left(\Omega_k \mid \bm{\theta}, I\right)
    },
\end{equation}
where $p\left(m_1, m_2 \mid \alpha\right)$ is our component mass
prior and $N_j$ is the number of posterior samples for the $j$-th detection.
The selection function $p(\bm{\mathrm{D}}_{j} \mid \Omega,I)$ is evaluated at
the time of detection of the $j$-th event with the event's \ac{PSD}.
Furthermore, in the final expression for the likelihood, the denominator of
Eq.~\eqref{eq:final_eventlikelihood} cancels with $\hat{N}_j$
(Eq.~\ref{eq:expected_detections_segment}) as
$p\left(\Omega_k \mid \bm{\theta}\right) \propto a_k$.

\section{Results}\label{sec:results}

\begin{figure}
  \centering
	\includegraphics[width=0.48\textwidth]{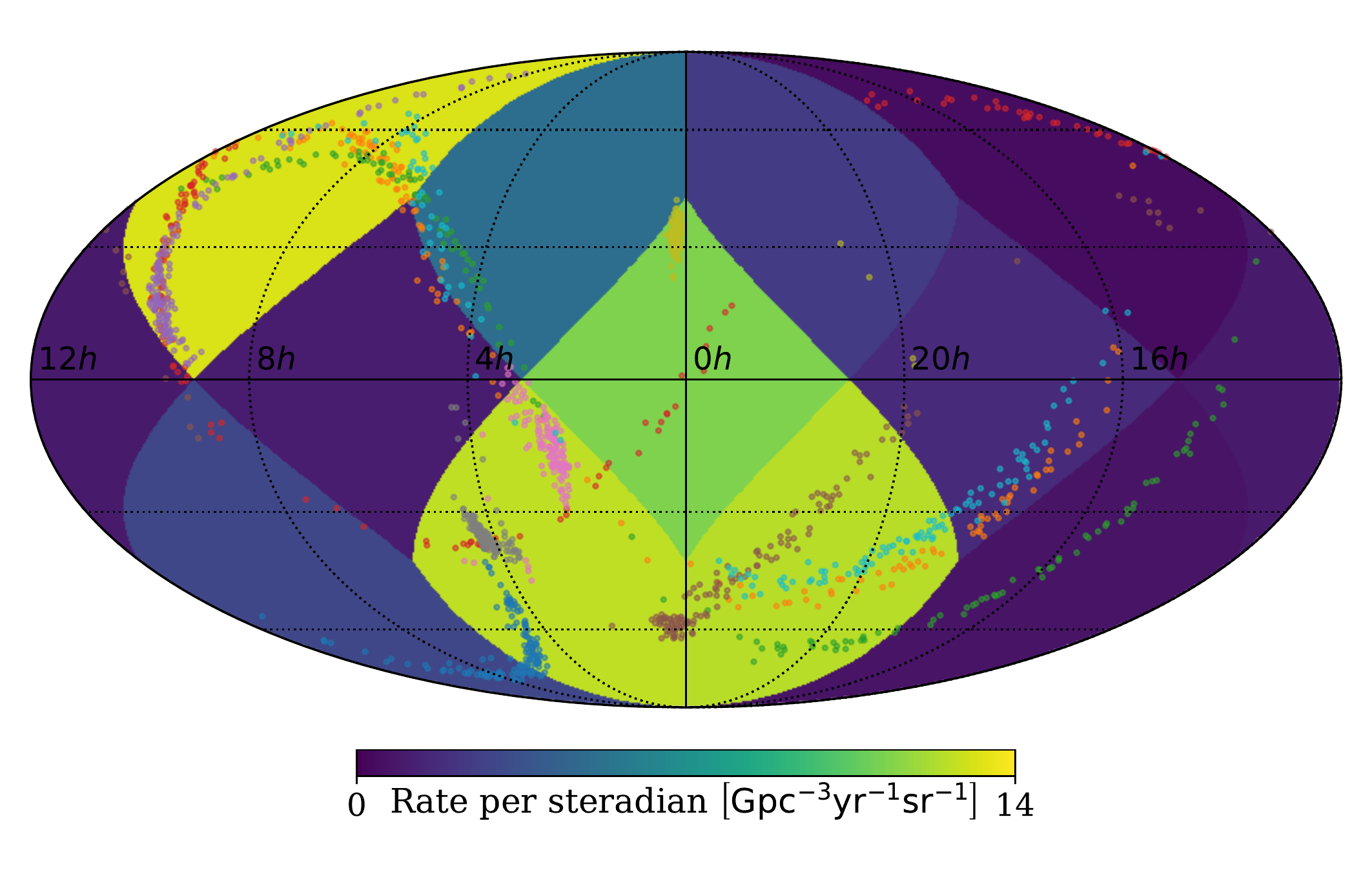}
    \caption{Example of a posterior sample for the anisotropic model, corresponding to
    the maximum likelihood probability. The GW event samples are rotated
    relative to the original coordinate system (Fig.~\ref{fig:eventLocalisation}).}
	\label{fig:post_example}
\end{figure}

The isotropic model is parametrised by a single pixel covering the
entire sky, making it invariant under the angular distribution of pixels and
independent of the Euler angles.
On the other hand, the anisotropic model spans a $15$-dimensional parameter space.
Each posterior sample in the anisotropic model contains a set of $12$ pixel
weights and its associated Euler angles describing the orientation of the HEALPix grid.
It follows then that in each sample the pixels correspond to different locations on
the sky, which is accounted for before averaging the pixel weights.
In Fig.~\ref{fig:post_example} we show the maximum likelihood sample
for the anisotropic model. To average the pixel weights,
the original $12$ pixels are split into a set of $49,152$ pixels which map the
samples onto a finer basis, while preserving their respective values. This
larger set is rotated and then averaged out as this corresponds to returning to
the original coordinate system, resulting in  a smoothing of the original
$12$-dimensional pixel basis and rendering the mean pixel weights on a
higher resolution map.

\begin{figure*}
	\centering
    \includegraphics[width=0.90\textwidth]{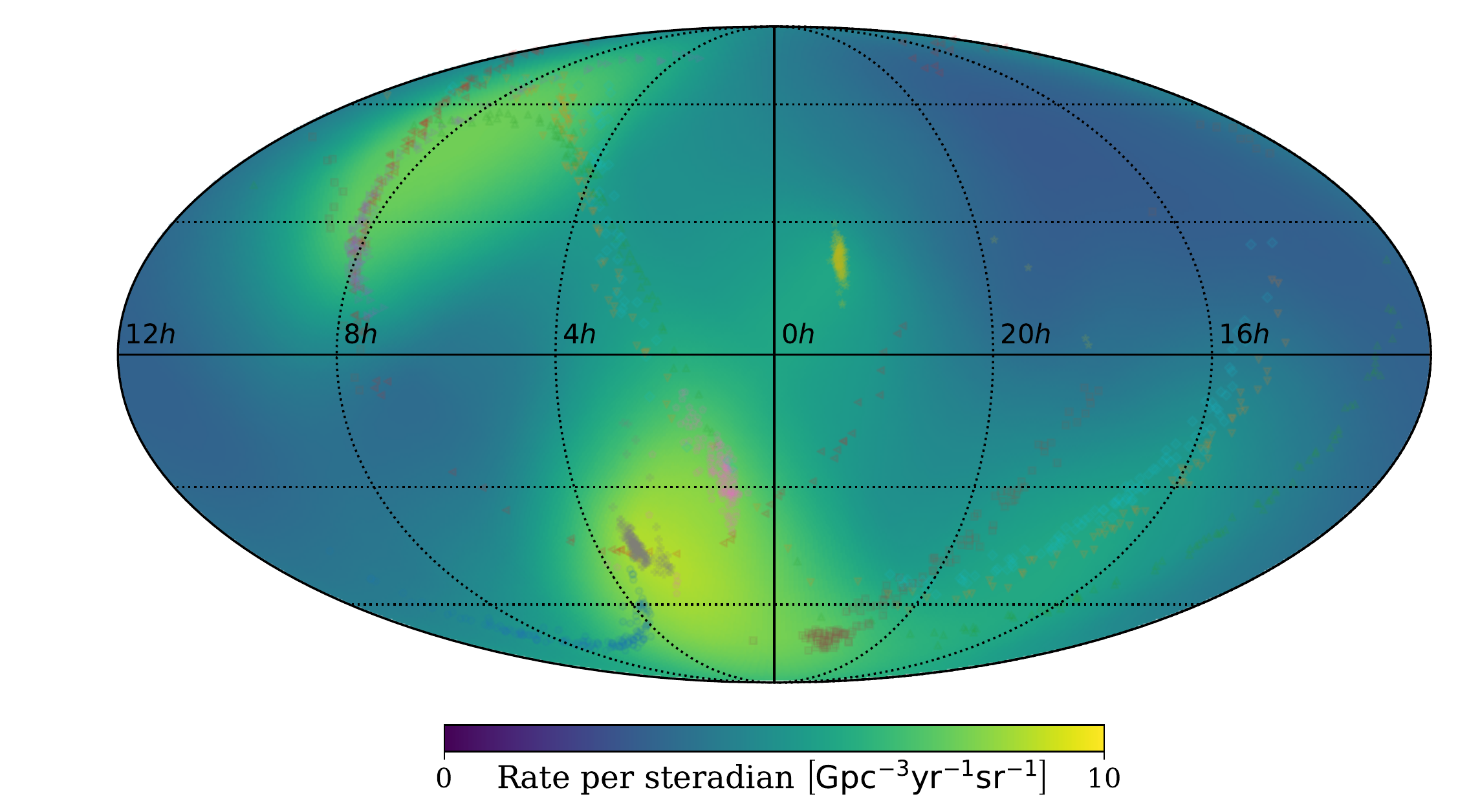}
    \caption{A mean map of pixel weights averaged over the different rotations of the
    posterior distributions such as Fig.~\ref{fig:post_example}. The individual
    samples were rendered on a smoother basis of $49,152$ pixels (the weights are
    defined per steradian, thus independent of the pixel size), rotated to a
    default frame of reference and averaged. The ratio of the maximum to the
    minimum rate density of the map is $\sim~3$.}
	\label{fig:smoothedMap}
\end{figure*}

The averaged pixel map of posterior samples for the anisotropic model is shown
in Fig.~\ref{fig:smoothedMap}. Two parts of the sky have above average values of
rate density, corresponding to the areas with the highest density of samples
(Fig.~\ref{fig:eventLocalisation}). However, the ratio of the maximum to the
minimum rate densities is $\sim{3}$, while the ratio of the maximum to the minimum
number of GW event posterior samples per pixel is $\sim{15}$
(Fig.~\ref{fig:eventLocalisation}), illustrating how the selection functions
differs between our posterior and a simple samples count.

\begin{table}
    \centering
    \begin{tabular}{l l}
    \hline\hline 
    Anisotropic model total rate ${R}_{\mathrm{ani}}$ &
    $60.5^{+33.4}_{-25.4}\,\mathrm{Gpc}^{-3}\mathrm{yr}^{-1}$\\[0.3ex]
    Isotropic model total rate ${R}_{\mathrm{iso}}$ &
    $58.9^{+42.4}_{-23.0}\,\mathrm{Gpc}^{-3}\mathrm{yr}^{-1}$\\[0.3ex]
    LVC total rate $R_{\mathrm{LVC}}$&
    $53.2^{+58.5}_{-28.8}\,\mathrm{Gpc}^{-3}\mathrm{yr}^{-1}$\\[0.3ex]
    Bayes factor $\mathcal{Z}_{\text{ISO}}/\mathcal{Z}_{\text{ANISO}}$ & $1.3:1$\\
    \hline\hline
    \end{tabular}
    \caption{Comparison of the total rate of our models with the LVC result
    and the Bayes factor ratio of the isotropic and anisotropic model.}
\label{table:results}
\end{table}

\begin{figure}
	\centering
	\includegraphics[width=0.45\textwidth]{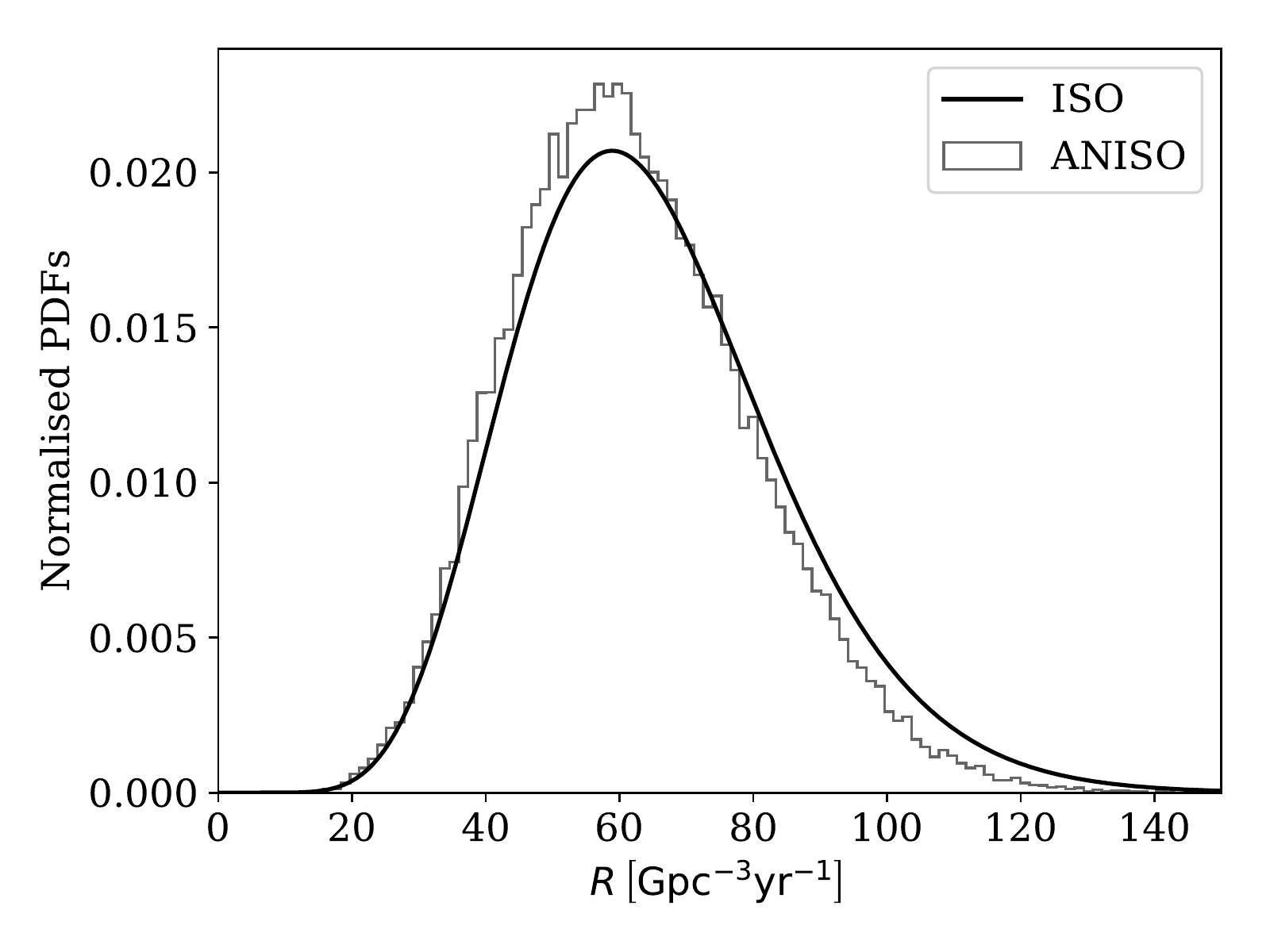}
    \caption{Posterior probability distributions on the rate $R$ of \ac{BBH} mergers,
	assuming the isotropic and anisotropic models. The isotropic model is
	consistent the LVC estimate of the total rate
	$53.2^{+58.5}_{-28.8}\,\mathrm{Gpc}^{-3}\mathrm{yr}^{-1}$ of \ac{BBH} mergers.}
	\label{fig:totalRateHist}
\end{figure}

Furthermore, we obtain estimates of the total rate $R$ (defined in
Eq.~\ref{eq:totalRate}), as in our analysis the merger rate is proportional
to the sum of the pixel weights. In Fig.~\ref{fig:totalRateHist} we show a histogram of
the total rate for the two models and quote our results with $90\%$ credibility
intervals in Table~\ref{table:results}.

The confidence intervals we obtain are tighter than the LVC estimate of the total
rate~\citep{O2populations}. This distinction can be attributed to a number of differing
assumptions between the LVC analysis and our own. These include, but are not
limited to the facts that our analysis makes different prior assumptions about
the black hole mass distribution, and that we do not marginalise over
uncertainty in the mass power law index and our model assumes a maximum total
mass. In addition, we have not used a realistic cosmological model for our
redshift distribution, we have considered only the inspiral component of the
signal waveform as opposed to the full inspiral, merger, and ringdown, and our
selection function is computed analytically and not empirically from simulated
signal injections into real non-Gaussian detector noise.

To determine whether the \ac{BBH} mergers are distributed
isotropically on the sky, we calculate a Bayes factor between the two
hypotheses (defined in Eq.~\ref{eq:evidence}). We arrive at a Bayes factor
$\mathcal{Z}_{\mathrm{ISO}} / \mathcal{Z}_{\mathrm{ANI}}$ of $1.3:1$, indicating
that the $10$ events from O1 and O2 runs show no statistically significant
preference for either model.

\begin{figure}
	\centering
	\includegraphics[width=0.45\textwidth]{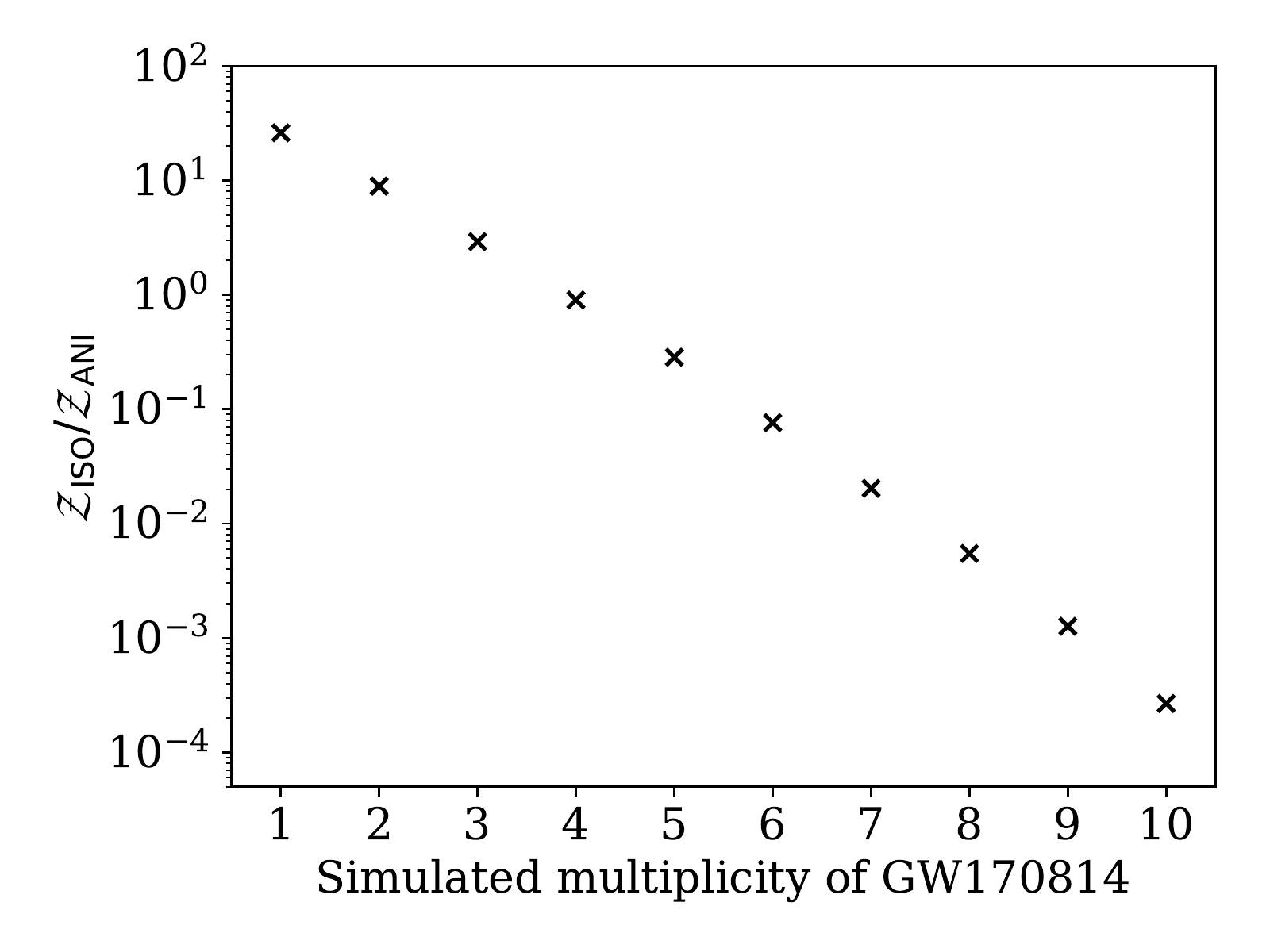}
    \caption{Bayes factor $\mathcal{Z}_{\mathrm{ISO}}/\mathcal{Z}_{\mathrm{ANISO}}$
    when all events
    from O1 and O2 runs are replaced with $n=1, 2, \ldots, 10$ copies of GW170814.
    Initially, for low values of $n$, the data prefers the isotropic model but
    as $n$ increases and more events are added into the same location the
    confidence in the anisotropic model exponentially grows.}
	\label{fig:bayes_factors}
\end{figure}

Lastly, we test our model by replacing all the observed events from O1 and O2 with
$n=1, 2, \ldots, 10$ copies of GW170814 and record the Bayes factors
$\mathcal{Z}_{\mathrm{ISO}}/\mathcal{Z}_{\mathrm{ANISO}}$ as a function of $n$. This is
shown in Fig.~\ref{fig:bayes_factors}. We observe an exponential dependence on $n$,
with isotropy being preferred for low values of $n$ but gradually showing strong
support for anisotropy as $n$ increases.

\section{Conclusion}\label{sec:conclusion}
We perform a Bayesian model selection, comparing an isotropic model to a
particular pixelated anisotropic model of the \ac{BBH} astrophysical merger
rate. We demonstrate that the $10$ detected \ac{BBH} events from O1 and O2
runs show comparable support for both models. This result was expected, the
underlying distribution of \ac{BBH} mergers is thought to be isotropic,
but with only $10$ events forming our dataset a strong statement about
isotropy should not be expected.

The anisotropic model is described with $15$ parameters which give a very large
parameter space for the model, most of which is incompatible with the data.
Since we made a decision to use a pixel basis, and the minimum number of pixels a
HEALpix map supports is $12$, it was not possible to easily switch to a lower
or higher dimensional basis. Thus, one of the future extensions of this project would
be to to perform this analysis in a spherical harmonics basis. A spherical
harmonic basis would allow for a lower dimensional parameter space,
for example, the simplest anisotropic spherical harmonics model would have $4$
weight parameters (polynomial degree $l_{\mathrm{max}}=1$). Such spherical
harmonic approaches could be readily used to model different anisotropy models,
providing us with an alternative representation with which to compare the
benchmark isotropic model. This approach is taken by \cite{Payne:2020pmc},
which also finds that the GWTC-1 data-set weakly favours the isotropic model,
although their analysis differs from ours in other respects.

In this analysis the prior \ac{PDF} on luminosity distance is consistent
with that of a simple, static Euclidean universe. As the interferometers are
being upgraded in terms of sensitivity it would be reasonable to include a
cosmological model of an expanding universe since some events are already at
significant redshifts. However, we do not expect this to change the result
dramatically, rather it should result in scaling the selection function and
providing us with better tools that could be used in further applications.

Finally, and potentially most importantly, at the time of writing, only ten
events (from O1 and O2) have been published by the LIGO-Virgo Collaboration. The
recently finished O3 run lived up to expectations and provided $\sim~60$
additional candidates and the O3 events are generally better
localised due to sensitivity improvements at the Virgo site~\citep{GraceDB}.
Such additional
data will only serve to strengthen the analysis presented here and we are eager
to use these events to compare our \ac{BBH} distribution models. In the
future we predict that this work can be used to significantly strengthen our
belief in the isotropic model over an anisotropic model as the number of
detections increases. Alternatively, the analysis can be interpreted as an
important probe for finding anisotropies, should they exist. In the limit of
a large number of detections, if \ac{BBH} sources trace the large scale structure
of the Universe we might expect that a general anisotropic model might be favoured
over isotropy. This scenario would allow us to meaningfully compare our results to
the distribution of visible matter in the Universe. 

\section*{Acknowledgements}

This work was supported by a Royal Astronomical Society summer undergraduate
research bursary. We thank Rachel Gray for useful discussions and we also thank
the anonymous referee for suggestions that have improved the work presented.
This research has made use of data, software and/or web tools obtained from the
Gravitational Wave Open Science Center (\url{https://www.gw-openscience.org}),
a service of LIGO Laboratory, the LIGO Scientific Collaboration and the Virgo
Collaboration. LIGO is funded by the U.S.  National Science Foundation. Virgo
is funded by the French Centre National de Recherche Scientifique (CNRS), the
Italian Istituto Nazionale della Fisica Nucleare (INFN) and the Dutch Nikhef,
with contributions by Polish and Hungarian institutes. JV was partially supported
by STFC grant ST/K005014/1 and JV and CM are supported by the Science and
Technology Research Council (grant No.~ST/~L000946/1).

\section*{Data Availability}
The data underlying this article will be shared on
reasonable request to the corresponding author.



\bibliographystyle{mnras}
\bibliography{ref} 

\begin{thebibliography}{}
\makeatletter
\relax
\def\mn@urlcharsother{\let\do\@makeother \do\$\do\&\do\#\do\^\do\_\do\%\do\~}
\def\mn@doi{\begingroup\mn@urlcharsother \@ifnextchar [ {\mn@doi@}
  {\mn@doi@[]}}
\def\mn@doi@[#1]#2{\def\@tempa{#1}\ifx\@tempa\@empty \href
  {http://dx.doi.org/#2} {doi:#2}\else \href {http://dx.doi.org/#2} {#1}\fi
  \endgroup}
\def\mn@eprint#1#2{\mn@eprint@#1:#2::\@nil}
\def\mn@eprint@arXiv#1{\href {http://arxiv.org/abs/#1} {{\tt arXiv:#1}}}
\def\mn@eprint@dblp#1{\href {http://dblp.uni-trier.de/rec/bibtex/#1.xml}
  {dblp:#1}}
\def\mn@eprint@#1:#2:#3:#4\@nil{\def\@tempa {#1}\def\@tempb {#2}\def\@tempc
  {#3}\ifx \@tempc \@empty \let \@tempc \@tempb \let \@tempb \@tempa \fi \ifx
  \@tempb \@empty \def\@tempb {arXiv}\fi \@ifundefined
  {mn@eprint@\@tempb}{\@tempb:\@tempc}{\expandafter \expandafter \csname
  mn@eprint@\@tempb\endcsname \expandafter{\@tempc}}}

\bibitem[\protect\citeauthoryear{Abbott et~al.}{Abbott et~al.}{2016a}]{O1BBH}
Abbott B.~P.,  et~al., 2016a, \mn@doi [{Phys. Rev. X}]
  {10.1103/PhysRevX.6.041015}, 6, 041015

\bibitem[\protect\citeauthoryear{Abbott et~al.}{Abbott
  et~al.}{2016b}]{GW150914}
Abbott B.~P.,  et~al., 2016b, \mn@doi [Phys. Rev. Lett.]
  {10.1103/PhysRevLett.116.061102}, 116, 061102

\bibitem[\protect\citeauthoryear{Abbott et~al.}{Abbott
  et~al.}{2016c}]{GW150914:RATES}
Abbott B.~P.,  et~al., 2016c, \mn@doi [Astrophys. J. Lett.]
  {10.3847/2041-8205/833/1/L1}, 833, L1

\bibitem[\protect\citeauthoryear{Abbott et~al.}{Abbott
  et~al.}{2017a}]{GW170104}
Abbott B.~P.,  et~al., 2017a, \mn@doi [Phys. Rev. Lett.]
  {10.1103/PhysRevLett.118.221101}, 118, 221101

\bibitem[\protect\citeauthoryear{Abbott et~al.}{Abbott
  et~al.}{2017b}]{GW170814}
Abbott B.~P.,  et~al., 2017b, \mn@doi [Phys. Rev. Lett.]
  {10.1103/PhysRevLett.119.141101}, 119, 141101

\bibitem[\protect\citeauthoryear{{Abbott} et~al.,}{{Abbott}
  et~al.}{2017c}]{GW170608}
{Abbott} B.~P.,  et~al., 2017c, \mn@doi [Ap. J. Lett.]
  {10.3847/2041-8213/aa9f0c}, \href
  {http://adsabs.harvard.edu/abs/2017ApJ...851L..35A} {851, L35}

\bibitem[\protect\citeauthoryear{Abbott et~al.}{Abbott
  et~al.}{2018}]{Aasi:2013wya}
Abbott B.~P.,  et~al., 2018, \mn@doi [Living Rev. Rel.]
  {10.1007/s41114-018-0012-9, 10.1007/lrr-2016-1}, 21, 3

\bibitem[\protect\citeauthoryear{Abbott et~al.}{Abbott et~al.}{2019a}]{H0paper}
Abbott B.~P.,  et~al., 2019a, preprint (\mn@eprint {arXiv} {1908.06060})

\bibitem[\protect\citeauthoryear{Abbott et~al.}{Abbott
  et~al.}{2019b}]{O1O2catalog}
Abbott B.~P.,  et~al., 2019b, \mn@doi [Phys. Rev.] {10.1103/PhysRevX.9.031040},
  X9, 031040

\bibitem[\protect\citeauthoryear{Abbott et~al.}{Abbott
  et~al.}{2019c}]{O2populations}
Abbott B.~P.,  et~al., 2019c, \mn@doi [Astrophys. J.]
  {10.3847/2041-8213/ab3800}, 882, L24

\bibitem[\protect\citeauthoryear{Acernese et~al.,}{Acernese
  et~al.}{2014}]{Virgo}
Acernese F.,  et~al., 2014, \mn@doi [Classical and Quantum Gravity]
  {10.1088/0264-9381/32/2/024001}, 32, 024001

\bibitem[\protect\citeauthoryear{Anderson, Brady, Creighton  \&
  Flanagan}{Anderson et~al.}{2001}]{Anderson:2000yy}
Anderson W.~G.,  Brady P.~R.,  Creighton J. D.~E.,   Flanagan E.~E.,  2001,
  \prd, 63, 042003

\bibitem[\protect\citeauthoryear{Chen, Essick, Vitale, Holz  \&
  Katsavounidis}{Chen et~al.}{2017}]{Chen:2016luc}
Chen H.-Y.,  Essick R.,  Vitale S.,  Holz D.~E.,   Katsavounidis E.,  2017,
  \mn@doi [Astrophys. J.] {10.3847/1538-4357/835/1/31}, 835, 31

\bibitem[\protect\citeauthoryear{{Daniel Sigg}}{{Daniel Sigg}}{2016a}]{PSDO1H1}
{Daniel Sigg} 2016a, {H1 Calibrated Sensitivity Spectra Oct 24 2015
  (Representative for O1)}, \url {https://dcc.ligo.org/LIGO-G1600150/public}

\bibitem[\protect\citeauthoryear{{Daniel Sigg}}{{Daniel Sigg}}{2016b}]{PSDO1L1}
{Daniel Sigg} 2016b, {L1 Calibrated Sensitivity Spectra Oct 24 2015
  (Representative for O1)}, \url {https://dcc.ligo.org/LIGO-G1600151/public}

\bibitem[\protect\citeauthoryear{{Dominik}, {Berti}, {O'Shaughnessy}, {Mandel}
  et~al.}{{Dominik} et~al.}{2015}]{2015ApJ...806..263D}
{Dominik} M.,  {Berti} E.,  {O'Shaughnessy} R.,  {Mandel} I.,   et~al., 2015,
  \mn@doi [\apj] {10.1088/0004-637X/806/2/263}, \href
  {http://adsabs.harvard.edu/abs/2015ApJ...806..263D} {806, 263}

\bibitem[\protect\citeauthoryear{{G{\'o}rski}, {Hivon}, {Banday}, {Wandelt},
  {Hansen}, {Reinecke}  \& {Bartelmann}}{{G{\'o}rski}
  et~al.}{2005}]{2005ApJ...622..759G}
{G{\'o}rski} K.~M.,  {Hivon} E.,  {Banday} A.~J.,  {Wandelt} B.~D.,  {Hansen}
  F.~K.,  {Reinecke} M.,   {Bartelmann} M.,  2005, \mn@doi [\apj]
  {10.1086/427976}, \href {http://adsabs.harvard.edu/abs/2005ApJ...622..759G}
  {622, 759}

\bibitem[\protect\citeauthoryear{Harry}{Harry}{2010}]{Harry:2010zz}
Harry G.~M.,  2010, \mn@doi [Class. Quant. Grav.]
  {10.1088/0264-9381/27/8/084006}, 27, 084006

\bibitem[\protect\citeauthoryear{{Horvath}, {Hakkila}  \& {Bagoly}}{{Horvath}
  et~al.}{2013}]{2013arXiv1311.1104H}
{Horvath} I.,  {Hakkila} J.,   {Bagoly} Z.,  2013, arXiv e-prints, \href
  {https://ui.adsabs.harvard.edu/abs/2013arXiv1311.1104H} {p. arXiv:1311.1104}

\bibitem[\protect\citeauthoryear{{LIGO Scientific Collaboration}}{{LIGO
  Scientific Collaboration}}{2015}]{aLIGO}
{LIGO Scientific Collaboration} 2015, Class. Quantum Grav., 32, 074001

\bibitem[\protect\citeauthoryear{{LIGO-Virgo Collaboration}}{{LIGO-Virgo
  Collaboration}}{2018}]{PSDO2}
{LIGO-Virgo Collaboration} 2018, {GWTC-1}, \url
  {https://dcc.ligo.org/LIGO-P1800374/public}

\bibitem[\protect\citeauthoryear{{LIGO-Virgo Collaboration}}{{LIGO-Virgo
  Collaboration}}{2019}]{PSDevent}
{LIGO-Virgo Collaboration} 2019, {Power Spectral Densities (PSD) release for
  GWTC-1}, \url {https://dcc.ligo.org/LIGO-P1900011/public}

\bibitem[\protect\citeauthoryear{{LIGO-Virgo Collaboration}}{{LIGO-Virgo
  Collaboration}}{2020}]{GraceDB}
{LIGO-Virgo Collaboration} 2020, {Gravitational-Wave Candidate Event Database},
  \url {https://gracedb.ligo.org/superevents/public/O3/}

\bibitem[\protect\citeauthoryear{Mandel, Farr  \& Gair}{Mandel
  et~al.}{2019}]{2018arXiv180902063M}
Mandel I.,  Farr W.~M.,   Gair J.~R.,  2019, \mn@doi [Mon. Not. Roy. Astron.
  Soc.] {10.1093/mnras/stz896}, 486, 1086

\bibitem[\protect\citeauthoryear{Ng, Vitale, Zimmerman, Chatziioannou, Gerosa
  \& Haster}{Ng et~al.}{2018}]{Ng:2018neg}
Ng K. K.~Y.,  Vitale S.,  Zimmerman A.,  Chatziioannou K.,  Gerosa D.,   Haster
  C.-J.,  2018, \mn@doi [Phys. Rev.] {10.1103/PhysRevD.98.083007}, D98, 083007

\bibitem[\protect\citeauthoryear{O'Shaughnessy, Vaishnav, Healy  \&
  Shoemaker}{O'Shaughnessy et~al.}{2010}]{PhysRevD.82.104006}
O'Shaughnessy R.,  Vaishnav B.,  Healy J.,   Shoemaker D.,  2010, \mn@doi
  [Phys. Rev. D] {10.1103/PhysRevD.82.104006}, 82, 104006

\bibitem[\protect\citeauthoryear{{Payne}, {Banagiri}, {Lasky}  \&
  {Thrane}}{{Payne} et~al.}{2020}]{Payne:2020pmc}
{Payne} E.,  {Banagiri} S.,  {Lasky} P.,   {Thrane} E.,  2020, arXiv e-prints,
  \href {https://ui.adsabs.harvard.edu/abs/2020arXiv200611957P} {p.
  arXiv:2006.11957}

\bibitem[\protect\citeauthoryear{{Saadeh}, {Feeney}, {Pontzen}, {Peiris}  \&
  {McEwen}}{{Saadeh} et~al.}{2016}]{2016PhRvL.117m1302S}
{Saadeh} D.,  {Feeney} S.~M.,  {Pontzen} A.,  {Peiris} H.~V.,   {McEwen} J.~D.,
   2016, \mn@doi [\prl] {10.1103/PhysRevLett.117.131302}, \href
  {https://ui.adsabs.harvard.edu/abs/2016PhRvL.117m1302S} {117, 131302}

\bibitem[\protect\citeauthoryear{Singer \& Price}{Singer \&
  Price}{2016}]{BAYESTAR}
Singer L.~P.,  Price L.,  2016, \mn@doi [\prd] {10.1103/PhysRevD.93.024013},
  93, 024013

\bibitem[\protect\citeauthoryear{Skilling}{Skilling}{2006}]{skilling2006}
Skilling J.,  2006, \mn@doi [Bayesian Anal.] {10.1214/06-BA127}, 1, 833

\bibitem[\protect\citeauthoryear{{Vallisneri}, {Kanner}, {Williams},
  {Weinstein}  \& {Stephens}}{{Vallisneri} et~al.}{2015}]{Vallisneri:2014vxa}
{Vallisneri} M.,  {Kanner} J.,  {Williams} R.,  {Weinstein} A.,   {Stephens}
  B.,  2015, in Journal of Physics Conference Series. p. 012021 (\mn@eprint
  {arXiv} {1410.4839}), \mn@doi{10.1088/1742-6596/610/1/012021}, \url
  {https://www.gw-openscience.org/about/}

\bibitem[\protect\citeauthoryear{Veitch et~al.}{Veitch
  et~al.}{2015}]{LALInference}
Veitch J.,  et~al., 2015, \mn@doi [\prd] {10.1103/PhysRevD.91.042003}, 91,
  042003

\bibitem[\protect\citeauthoryear{Veitch, {Del Pozzo}, Cody, Pitkin  \&
  ed1d1a8d}{Veitch et~al.}{2017}]{john_veitch_2017_835874}
Veitch J.,  {Del Pozzo} W.,  Cody Pitkin M.,   ed1d1a8d 2017,
  \url{http://www.github.com/johnveitch/cpnest},
  \mn@doi{10.5281/zenodo.835874}, \url {https://doi.org/10.5281/zenodo.835874}

\bibitem[\protect\citeauthoryear{{Virgo Collaboration}}{{Virgo
  Collaboration}}{2014}]{AdVirgo}
{Virgo Collaboration} 2014, \mn@doi [Class. Quantum Grav.]
  {10.1088/0264-9381/32/2/024001}, 32, 024001

\makeatother
\end{thebibliography}


\bsp	
\label{lastpage}
\end{document}